\documentclass[aps,prl,twocolumn,superscriptaddress,amsmath,amssymb]{revtex4-2}

\usepackage{mathtools}
\usepackage{braket}
\usepackage{amsthm} % Added for Supplemental Material theorems
\usepackage[utf8]{inputenc}
\usepackage[colorlinks=true,linkcolor=blue,citecolor=blue,urlcolor=blue]{hyperref}

\newcommand{\Tr}{\mathrm{Tr}}
\newcommand{\I}{\mathbb{I}} % Identity operator

\newcommand{\Var}{\mathrm{Var}}
\newcommand{\Dbath}{D_{\mathrm{bath}}}

\newtheorem{lemma}{Lemma}

\newtheorem{theorem}{Theorem}
\newtheorem{corollary}{Corollary}
\theoremstyle{remark}

\DeclareMathOperator{\csch}{csch}

\begin{document}

\title{Matrix Thermodynamic Uncertainty Relation for Non-Abelian Charge Transport}

\author{Domingos S.\ P.\ Salazar}
\affiliation{Universidade Federal Rural de Pernambuco, Departamento de F\'isica, Recife, PE, Brazil}

\begin{abstract}
Thermodynamic uncertainty relations (TURs) bound the precision of currents by entropy production, but quantum transport of noncommuting (non-Abelian) charges challenges standard formulations because different charge components cannot be monitored within a single classical frame. We derive a process-level \emph{matrix} TUR starting from the operational entropy production $\Sigma = D(\rho'_{SE}\|\rho'_S\!\otimes\!\rho_E)$. Isolating the experimentally accessible bath divergence $\Dbath=D(\rho'_E\|\rho_E)$, we prove a fully nonlinear, \emph{saturable} lower bound valid for arbitrary current vectors $\Delta q$:
$\Dbath \ge B(\Delta q,V,V')$, where the bound depends only on the transported-charge signal $\Delta q$ and the pre/post collision covariance matrices $V$ and $V'$. In the small-fluctuation regime $\Dbath\geq\frac12\,\Delta q^{\mathsf T}V^{-1}\Delta q+O(\|\Delta q\|^4)$, while beyond linear response it remains accurate. Numerical strong-coupling qubit collisions illustrate the bound and demonstrate near-saturation across broad parameter ranges using only local measurements on the bath probe.
\end{abstract}

\maketitle

The thermodynamic uncertainty relation (TUR) expresses a universal tradeoff between
current fluctuations and entropy production in nonequilibrium steady states \cite{Barato2015,Gingrich2016,Horowitz2020Rev,Seifert2019} and later adapted to more general classical and quantum settings \cite{Koyuk2019,LandiReview2021,Liu2019,VanVu2023,Miller2021,Pires2021,Guarnieri2019,Hasegawa2019a,VanVu2019,VanVu2021,Hasegawa2021,Yan2019,MacIeszczak2018,Carollo2019,Ishida2025,Prech2025,Brandner2025,Moreira2025}.
Recent work has highlighted how quantum thermodynamics becomes qualitatively subtler in the presence of
multiple conserved quantities (``charges'') that need not commute \cite{Guryanova2016,MajidyHalpernRMP2023}.
Such \emph{noncommuting} (non-Abelian) charges are central to SU(2) spin transport and generalized Gibbs
ensembles, and can lead to dissipation/coherence effects absent in commuting (Abelian) settings
\cite{HalpernMajidy2022,Twesh2024}.

Classical matrix TURs \cite{Timpanaro2019B,Dechant2018} typically assume access to a \emph{joint} classical record of all currents, so that covariances are defined within a single measurement frame.
    In contrast, for multiple \emph{noncommuting} charges there is no single measurement setting that yields their joint statistics without additional backaction \cite{Heinosaari2016,Busch2013,Guhne2023,Strasberg2021new}.
Here we circumvent this obstruction by bounding dissipation using \emph{separate} measurement settings on the bath probe: only the first moments $\Delta q$ and the corresponding pre/post covariances $V,V'$ are required.
Our bound is kinematic and applies beyond weak coupling, without invoking time-reversal symmetry or linear-response assumptions.

A core conceptual obstacle is that ``entropy production'' in open quantum dynamics is not naturally
a divergence between a state and a time-reversed state. A particularly clean microscopic definition
is the process-level entropy production (EP) \cite{Esposito2010a}
\begin{equation}
\label{eq:EP_def}
\Sigma
:= D\!\left(\rho'_{SE}\,\big\|\,\rho'_S\otimes\rho_E\right),
\end{equation}
where $D(\rho\|\sigma)=\Tr(\rho(\ln\rho-\ln\sigma))$ is the quantum relative entropy and the initial product state $\rho_S\otimes\rho_E$ evolves unitarily to $\rho'_{SE}$ and
$\rho'_S:=\Tr_E\rho'_{SE}$. Eq.~\eqref{eq:EP_def} equals the final mutual information $I(S\!:\!E):=D(\rho_{SE}'\|\rho_S'\otimes \rho_E')$ plus the bath divergence $\Dbath:=D(\rho'_E\|\rho_E)$,
\begin{equation}
\label{eq:EP_split}
\Sigma = I(S\!:\!E) + \Dbath \ \ge\ \Dbath.
\end{equation}
This definition is well-suited to non-Abelian transport because it requires no choice of reversed dynamics.

Our first message is that the familiar Onsager (linear-response) structure \cite{Onsager1931,Kubo1957,Andrieux2004,Saito2008,Salazar2020a,ManzanoParrondoLandi2022} emerges as a small-fluctuation limit of
a fully nonlinear, \emph{matrix} TUR.  Consider a vector of transported charges
$q=\big(q_u\big)_u$ measured on $E$ (e.g.\ spin components), with mean change
\begin{equation}
\Delta q_u := \Tr\!\left[(\rho'_E-\rho_E)\,Q_u\right], 
\qquad 
\Delta q := (\Delta q_u)_{u=1}^m .
\label{eq:Deltaq_def}
\end{equation}
When $[U,\,Q_u^S+Q_u^E]=0$, the quantity $\Delta q_u$ is a bona fide transported charge (exchange between system and probe); otherwise it is an observable displacement on the probe. The bound below is kinematic and applies in both cases. When $[U,\,Q_u^S+Q_u^E]=0$, this coincides (up to sign) with the change of the corresponding charge on $S$ \cite{LandiReview2021}.

Let $V$ and $V'$ denote the charge covariance matrices evaluated on $\rho_E$ and $\rho'_E$, respectively
(precise definition given below).  In the small-fluctuation regime,
the $\Dbath$ admits the expansion in an Onsager quadratic form,
\begin{equation}
\label{eq:Onsager_limit_intro}
\Dbath \geq \frac12\,\Delta q^{\mathsf T}V^{-1}\Delta q + O(\|\Delta q\|^4),
\end{equation}
recovering the classical TUR scaling while retaining a genuinely quantum (non-Abelian) covariance geometry
through $V$.

Our main result, however, holds \emph{far beyond linear response} and for any current vector $\Delta q$, including the case where $[Q_i,Q_j]\neq 0$.
We prove an explicit, saturable lower bound:
\begin{equation}
\label{eq:main_chain_intro}
\Sigma \ \ge\ \Dbath
\ \ge\ 
\int_{0}^{1} \frac{\lambda\, s_\lambda}{1+\lambda(1-\lambda)\, s_\lambda}\, d\lambda,
\end{equation}
where the scalar $s_\lambda$ is a quadratic form built from the same three experimentally accessible ingredients
$(\Delta q,V,V')$,
\begin{equation}
\label{eq:s_lambda_intro}
s_\lambda := \Delta q^{\mathsf T}\, V_\lambda^{-1}\, \Delta q,
\qquad
V_\lambda := (1-\lambda)\,V' + \lambda\,V.
\end{equation}
Equations~\eqref{eq:main_chain_intro}--\eqref{eq:s_lambda_intro} constitute a \emph{matrix TUR} for transported,
possibly noncommuting charges: the full nonlinear bound reduces to the Onsager form
\eqref{eq:Onsager_limit_intro} for small fluctuations, while remaining valid for arbitrary fluctuations and often near-saturated in our simulations.

%============================================================
{\bf Formalism}
%============================================================
We consider one collision between a system $S$ and a fresh bath probe $E$ prepared in $\rho_E$ \cite{Scarani2002,Ciccarello2022},
\begin{equation}
\rho'_{SE}=U_{SE}(\rho_S\otimes\rho_E)U_{SE}^\dagger,
\end{equation}
with $\rho'_S=\Tr_E\rho'_{SE}$ and $\rho'_E=\Tr_S\rho'_{SE}$. The process-level entropy production $\Sigma$ is defined in Eq.~\eqref{eq:EP_def} and splits as in
Eq.~\eqref{eq:EP_split}.

We use the $m$-component current vector $\Delta q$ and the covariance matrices $V$ and $V'$ defined as
\begin{equation}
    V_{uv}:=\frac{1}{2}\Tr\!\left[\rho_E\{\Delta Q_u,\Delta Q_v\}\right],
    \label{FormalismV1}
    \end{equation}
    \begin{equation}
    V'_{uv}:=\frac{1}{2}\Tr\!\left[\rho'_E\{\Delta Q'_u,\Delta Q'_v\}\right],
    \label{FormalismV2}
\end{equation}
where $\Delta Q_u=Q_u-\Tr(\rho_E Q_u)\I$ and $\Delta Q'_u=Q_u-\Tr(\rho'_E Q_u)\I$.
Define the $\lambda$-interpolated covariance $V_\lambda$ and the associated scalar $s_\lambda$ (\ref{eq:s_lambda_intro}),
with Moore--Penrose inverses understood when needed.
Our main result (\ref{eq:main_chain_intro}) is
\begin{align}
\label{eq:formalism_main_chain}
\Dbath
\ \ge\ B(\Delta q,V,V')
:= \int_0^1 d\lambda\;
\frac{\lambda\, s_\lambda}{1+\lambda(1-\lambda)\,s_\lambda}.
\end{align}
This is a \emph{matrix TUR} valid for arbitrary currents $\Delta q$ (not restricted to small fluctuations). The proof is based on two steps:

\emph{(i) Integral representation in terms of $\chi^2_\lambda$.}
The bath relative entropy admits an exact integral representation \cite{Salazar2025}
\begin{equation}
\label{eq:formalism_KL_chi2}
D(\rho'_E\|\rho_E)=\int_0^1 d\lambda\;\lambda\,
\chi^2_\lambda(\rho'_E\|\rho_E),
\end{equation}
where $\chi^2_\lambda(\cdot\|\cdot)$ is the quantum $\chi^2_\lambda$-divergence
indexed by $\lambda$ (see SM for the explicit definition and basic properties) that is the basic atomic divergence for the scalar TUR. 

\emph{(ii) From a scalar witness bound to the matrix form.}
For any direction $u\in\mathbb{R}^m$, consider the scalar current
$J_u:=u^{\mathsf T}\Delta q$ and the scalar variances
$u^{\mathsf T}Vu$ and $u^{\mathsf T}V'u$.
Using Cauchy--Schwarz/Chapman--Robbins in the $\chi^2_\lambda$-geometry with the witness observable
$u\!\cdot\! q$ yields, for each fixed $\lambda$,
\begin{equation}
\label{eq:formalism_scalar_TUR}
\chi^2_\lambda(\rho'_E\|\rho_E)
\ \ge\
\frac{(J_u)^2}{u^{\mathsf T}V_\lambda u}\,
\frac{1}{1+\lambda(1-\lambda)\,\frac{(J_u)^2}{u^{\mathsf T}V_\lambda u}},
\end{equation}
where the prefactor is the familiar signal-to-noise ratio and the second factor is the nonlinear tightening
responsible for the saturable integrand in Eq.~\eqref{eq:formalism_main_chain}.
Optimizing Eq.~\eqref{eq:formalism_scalar_TUR} over $u$ at fixed $\lambda$ gives the closed matrix expression
in terms of $s_\lambda$ in Eq.~\eqref{eq:s_lambda_intro}.
Substituting the optimized bound into Eq.~\eqref{eq:formalism_KL_chi2} yields Eq.~\eqref{eq:formalism_main_chain} (details in the SM).

%============================================================
{\bf Discussion}
%===========================================================
{\it Onsager limit.}
In a near-fixed-point regime one has $V'\approx V$ (hence $V_\lambda\approx V$ and $s_\lambda\approx \Delta q^{\mathsf T}V^{-1}\Delta q$),
so Eq.~\eqref{eq:formalism_main_chain} reduces to the quadratic Onsager form stated in Eq.~\eqref{eq:Onsager_limit_intro},
while remaining valid far beyond linear response and often near-saturated in our numerics.

\paragraph{Symmetric-covariance regime $V'=V$.}
In the special case where the covariance is unchanged by the collision, $V'=V$, the interpolation
\eqref{eq:s_lambda_intro} becomes $\lambda$-independent, $V_\lambda=V$, and hence $s_\lambda$ is constant:
\begin{equation}
\label{eq:s_const}
s_\lambda \equiv s := \Delta q^{\mathsf T}V^{-1}\Delta q .
\end{equation}
The bound \eqref{eq:formalism_main_chain} then collapses to a one-parameter function which admits the closed form
\begin{equation}
\label{eq:F_closed}
\Dbath \ \ge F(s)
= 2\sqrt{\frac{s}{s+4}}\;
\operatorname{artanh}\!\left(\sqrt{\frac{s}{s+4}}\right).
\end{equation}
Physically, $V'=V$ is feasible whenever the probe state is (approximately) stationary under the collision at the level of
second moments of the measured charges. This occurs, for example, (a) in a near-fixed-point/linear-response regime where
$\rho_E'\approx\rho_E$ so that covariances change only at higher order in the current, (b) in weak-coupling or large-probe limits
where a single collision perturbs the probe by $O(1/N)$ and its covariance self-averages. In these regimes the full matrix bound reduces to the universal curve $F(s)$, while deviations from $V'=V$ quantify how far the
process operates from this symmetric-covariance idealization.

\paragraph{Matrix TUR in the symmetric-covariance regime ($V'=V$).}
We have seen above that when $V'=V$ the bound reduces to $\Dbath\ge F(s)$ with $s=\Delta q^{\mathsf T}V^{-1}\Delta q$
[Eqs.~\eqref{eq:s_const}--\eqref{eq:F_closed}]. Since $F(s)$ is strictly increasing for $s\ge 0$, it can be inverted to yield
\begin{equation}
\label{eq:s_le_G}
s \ \le\ G(\Dbath),
\qquad
G:=F^{-1}.
\end{equation}
Introducing the same auxiliary inverse function as in the classic (commuting) case ~\cite{Timpanaro2019},
namely $g(x)$ defined as the inverse of $x\mapsto x\tanh x$, one convenient closed form is obtained by writing
$\Dbath=2y\tanh y$ with $y=g(\Dbath/2)$, which gives $G(D)
=4\sinh^2\!\left(g\!\left(\frac{D}{2}\right)\right)$. We define
\begin{equation}
\label{eq:f_def_after_F}
f(D):=\frac{1}{G(D)}
=\frac{1}{4}\,\csch^2\!\left(g\!\left(\frac{D}{2}\right)\right).
\end{equation}
Then the scalar inequality \eqref{eq:s_le_G} is equivalent to the \emph{Matrix Symmetric Quantum TUR}, which is our second main result valid for $V=V'$,
\begin{equation}
\label{eq:matrix_TUR_sym_cov}
V - f(\Dbath)\,\Delta q\,\Delta q^{\mathsf T}\ \succeq\ 0,
\end{equation}
i.e., the left-hand side is positive semidefinite. Equation~\eqref{eq:matrix_TUR_sym_cov} has the same \emph{formal} structure a classical matrix exchange TUR
(e.g.\ Ref.~\cite{Timpanaro2019}), but its content is intrinsically quantum. When the charges commute, we recover the classical matrix TUR, where $V$ reduces to the classical covariance matrix.

Here the current is a \emph{vector of expectation-value changes of noncommuting charges} (for instance, in our simulation,
$Q_X=\sigma_X/2$ and $Q_Z=\sigma_Z/2$ with $[Q_X,Q_Z]\neq 0$), so there is no underlying joint distribution of
simultaneously measurable observables from which $V$ could be interpreted as a classical covariance matrix.
Rather, $V$ is a symmetrized quantum covariance geometry, reconstructible operationally from repeated preparations
with different measurement settings.
The symmetric-covariance condition $V'=V$ should therefore be viewed as an idealized regime in which the collision
preserves second moments of the chosen non-Abelian charges as already discussed.

\begin{figure}[t]
  \centering
  \includegraphics[width=\linewidth]{Fig1a.png}
  \caption{
  \textbf{Bath relative entropy bound from microscopic simulations.}
  For each random instance ($n=10^3$) of a system--bath interaction (cartoon inset), we simulate the reduced bath state
  $\rho_E\!\to\!\rho_E'$ and compute the \emph{bath relative entropy}
  $\Dbath:=D(\rho_E'\| \rho_E)$ (vertical axis).
  The horizontal axis shows our bound $B(\Delta q,V,V')$, built from three empirically accessible ingredients:
  the vector of current variations $\Delta q$ and the associated covariance matrices $V$ and $V'$ (before/after the interaction) color-coded by the relative covariance drift
  $\|V'-V\|/\|V\|$ after the unitary.
  The dashed diagonal indicates \emph{saturation}, $\Dbath=B(\Delta q,V,V')$. Inset: symmetric-covariance approximation $B\simeq F(s)$ as a function of
$s=\delta q^{\mathsf T}V^{-1}\delta q$ (black), where $\delta q\equiv\Delta q$ denotes the small-current regime.
The small-$s$ expansion $F(s)=s/2-s^2/12+O(s^3)$ (red dashed) captures the initial curvature, while deviations
appear for larger currents ($s\gg 1$), as expected beyond linear response.}

  \label{fig:Fig1a}
\end{figure}

\begin{figure}[t]
    \centering
    \includegraphics[width=\linewidth]{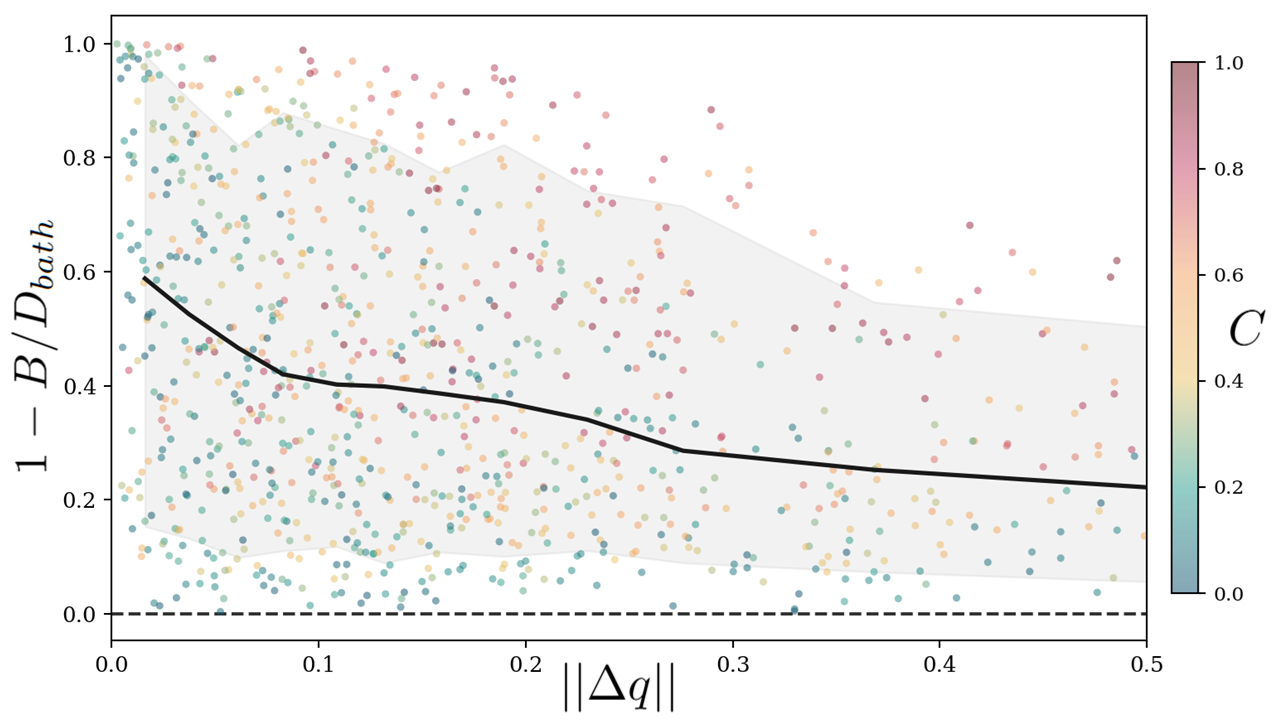}
    \caption{%
    \textbf{Near-saturation of the matrix TUR bound.}
    Scatter of the relative slack $1-B/\Dbath$ versus the current magnitude $\|\Delta q\|$ for the qubit collision model.
    Here $\Dbath$ is the bath relative entropy (entropy production contribution associated with the bath state change), and $B\equiv B(\Delta q,V,V')$ is the predicted lower bound from the full (non-Abelian) TUR.
    The dashed line at $0$ marks exact saturation ($B=\Dbath$).
    Point colors encode the noncommutativity indicator $C\in[0,1]$ used in the main text.
    The solid black curve is a binned running average (guide to the eye), and the light-gray band indicates the corresponding spread within bins.%
    }
    \label{fig:Fig2_qubit_saturation}
\end{figure}

% ============================================================
{\bf Application: strong-coupling collision simulation}
% ===================================================
Collision/repeated-interaction architectures—sequential strong system–ancilla couplings with ancilla re-preparation—are now experimentally routine across photonic platforms, trapped-ion open-system simulators, superconducting circuits with fast ancilla reset, and even gate-based quantum hardware \cite{Uriri2020,Visal2025,Andersen2019,Cattaneo2023,Haroche2013,Cuevas2019}. We model a \emph{single-shot} nonequilibrium transport collision simulation. here a system qubit $S$
collides once with a fresh probe qubit $E$ (the ``bath probe'').
Each run starts from a product preparation $\rho_S\otimes\rho_E$, undergoes a tunable
\emph{strong} entangling interaction $U_{SE}$, and ends with reduced output states
$\rho'_S$ and $\rho'_E$.
The process-level entropy production is $\Sigma$ in Eq.~\eqref{eq:EP_def}, and the directly accessible bath relative entropy is $\Dbath$ [Eq.~\eqref{eq:EP_split}].

The probe qubit $E$ is prepared in a full-rank state, conveniently parameterized by a Bloch vector,
\begin{equation}
\rho_E=\frac{1}{2}\Big(\I + r\,\hat n\cdot \vec\sigma\Big),
\qquad 0<r<1,\ \ \hat n\in S^2,
\end{equation}
so $\rho_E$ can be a polarized ``thermal-like'' qubit along direction $\hat n$ with polarization $r$. A \emph{fresh} copy of $E$ is used for each run (collision model).

The system qubit is prepared independently.
To scan both near-equilibrium and far-from-equilibrium regimes, we use either:
(i) a small coherent rotation away from $\rho_E$ (near-fixed-point regime), or
(ii) a generic state with different Bloch direction (far-from-equilibrium).
In all cases the protocol uses only the prepared $\rho_S$ and $\rho_E$ and the measured output $\rho'_E$. The collision is implemented by a tunable partial SWAP, realizable for instance via an exchange/Heisenberg
interaction,
\begin{equation}
U_{\mathrm{swap}}(\phi)=e^{-i\phi\,\mathrm{SWAP}}
=\cos(\phi)\,\I_{SE}-i\sin(\phi)\,\mathrm{SWAP},
\end{equation}
$\phi\in(0,\pi/2]$ (details in the SM). We take the transported charges to be two spin components on the probe qubit, chosen explicitly as
\begin{equation}
q=\big(q_1,q_2\big),
\quad
q_1 \equiv Q_X := \frac{1}{2}\,\sigma_X,
\quad
q_2 \equiv Q_Z := \frac{1}{2}\,\sigma_Z,
\end{equation}
so that $m=2$ and $[Q_X,Q_Z]\neq 0$.
This choice provides the minimal non-Abelian setting while keeping all quantities directly accessible in the simulation. We define the measured current on the bath probe as the change in charge expectation values:
\begin{equation}
\Delta q_u := \Tr\!\big[(\rho'_E-\rho_E)\,Q_u\big],
\qquad
\Delta q=\big(\Delta q_u\big)_{u=1}^m\in\mathbb{R}^m.
\end{equation}
Each $\Delta q_u$ is obtained from repeated runs by measuring $Q_u$ on $E$
before the collision (state $\rho_E$) and after the collision (state $\rho'_E$).
Because the $Q_u$ do not commute, different components are estimated from separate measurement settings
(rotating the measurement basis between runs).

We also estimate the charge covariances on $E$ before and after the collision using (\ref{FormalismV1}-\ref{FormalismV2}). Operationally, $V$ and $V'$ are reconstructed from the same repeated measurements used to estimate means:
one collects outcome statistics for each measurement basis and forms the empirical covariance matrix
(again requiring multiple settings when the charges do not commute).
We use the Frobenius (Hilbert--Schmidt) norm $\|A\|
:= \sqrt{\Tr(A^\dagger A)}$,
so the relative covariance drift is reported as $\|V-V'\|/\|V\|$. We also evaluate $\Dbath=D(\rho'_E\|\rho_E)$,
which only involves states of the probe qubit.
Experimentally, $\Dbath$ can be obtained by single-qubit tomography (or any equivalent
state-reconstruction method such as classical shadows), followed by classical post-processing to compute the
quantum relative entropy. From $(\Delta q,V,V')$ we form
$V_\lambda$ and $s_\lambda$ from (\ref{eq:s_lambda_intro})
and the bound
\begin{equation}
\Dbath\ge B(\Delta q,V,V'),
\end{equation}
where $B$ is given by (\ref{eq:formalism_main_chain}). We show the results in ~Fig.~\ref{fig:Fig1a}. All inputs to $B(\Delta q,V,V')$ are \emph{experimentally accessible} from repeated measurements on the bath probe
before and after the collision; no time-reversed dynamics, no weak-coupling assumption, and no joint
$SE$ tomography is required. When the collision perturbs the bath only weakly, one finds $V'\approx V$, so $V_\lambda\approx V$ and the bound
reduces to the Onsager quadratic form $\Delta q^{\mathsf T}V^{-1}\Delta q$ (\ref{eq:Onsager_limit_intro}).
At strong coupling, $V'$ can differ substantially from $V$, and the full nonlinear integral bound remains valid. 

In Fig.~\ref{fig:Fig2_qubit_saturation}, we quantify the state-dependent incompatibility of the two measured charges $(Q_1,Q_2)$ by the
\emph{Robertson incompatibility ratio}
\begin{equation}
\label{eq:C_robertson}
C \;\equiv\; \frac{\Big|\Im\,\Tr\!\big(\rho_E\,[Q_1,Q_2]\big)\Big|}{2\sqrt{V_{11}\,V_{22}}+\varepsilon},
\qquad \varepsilon = 10^{-15},
\end{equation}
where $V$ is the symmetrized covariance matrix evaluated on $\rho_E$.
Since $\langle[Q_1,Q_2]\rangle_{\rho_E}=\Tr(\rho_E[Q_1,Q_2])$ is purely imaginary for Hermitian $Q_1,Q_2$,
Eq.~\eqref{eq:C_robertson} is equivalently
\begin{equation}
C \;=\; \frac{\big|\Tr(\rho_E\, i[Q_1,Q_2])\big|}{2\sqrt{\Var_{\rho_E}(Q_1)\Var_{\rho_E}(Q_2)}+\varepsilon},
\end{equation}
where $\Var_{\rho_E}(Q_u)=V_{uu}$.
By the Robertson uncertainty relation, $|\langle[Q_1,Q_2]\rangle_{\rho_E}|\le 2\sqrt{\Var_{\rho_E}(Q_1)\Var_{\rho_E}(Q_2)}$,
hence $0\le C\le 1$ (up to the numerical regulator $\varepsilon$) \cite{Guhne2023}. While the drift $V\mapsto V'$ can occur even in commuting (Abelian) settings, it becomes especially relevant when the charges are incompatible:
For noncommuting charges there is no single joint measurement frame; the indicator $C\in[0,1]$ quantifies
\emph{state-dependent} incompatibility of the chosen pair in $\rho_E$, and hence how strongly multi-setting
estimation is forced.
In this sense, $C$ provides a non-Abelian witness (measurement-frame obstruction), and $V\mapsto V'$ quantifies the associated covariance geometry.
Our bound applies in this incompatible-charge setting and is often near-saturated in our numerics.

%============================================================
{\bf Outlook and Conclusions}
%============================================================
We developed a \emph{matrix} thermodynamic uncertainty relation for transport of \emph{noncommuting}
charges, starting from the process-level entropy production
$\Sigma=D(\rho'_{SE}\|\rho'_S\!\otimes\!\rho_E)$ and isolating the experimentally accessible bath
relative entropy $\Dbath=D(\rho'_E\|\rho_E)$.
Our central result is a fully nonlinear, \emph{saturable} bound
\begin{equation}
\Dbath\ \ge\ B(\Delta q,V,V') ,
\end{equation}
whose only inputs are the probe-side current vector $\Delta q$ and the pre/post charge-covariance matrices
$(V,V')$.
In the small-current regime this reduces to the Onsager quadratic form, while beyond linear response the
\emph{covariance drift} $V\!\to\!V'$ quantifies the breakdown of any single-covariance description.
Crucially, non-Abelianity enters operationally through incompatibility: for noncommuting charges there is no
single joint measurement frame, so one must combine statistics from distinct settings (captured in our
incompatibility indicator $C$).
Our witness resolves this core obstruction: it certifies dissipation without weak-coupling, Markovian, or
time-reversal assumptions, using only first/second moments accessible from local measurements (or
tomography/shadows) on the bath probe alone. In our collision-model numerics the bound is often near-saturated
across broad parameter ranges, including strongly coupled regimes.

Several directions now look ripe.
(i) Extend the one-shot witness to multi-collision and continuous-time dynamics, making explicit how the
resulting non-Abelian covariance geometry governs finite-time precision--dissipation tradeoffs and mixing.
(ii) Characterize equality and near-equality conditions, turning the TUR into a \emph{design principle} for
optimal non-Abelian transport protocols and minimal-dissipation gates.
(iii) Deploy the framework in platforms where incompatible components are natural and measurable (e.g.\ SU(2)
spin transport, cold atoms, and superconducting-qubit collision experiments), where both $C$ and the drift
$V\!\to\!V'$ provide directly testable signatures of genuinely quantum nonequilibrium transport.
Finally, the same logic extends beyond KL via the $\chi^2_\lambda$ decomposition of Petz $f$-divergences,
suggesting a family of tunable ``geometric'' TURs matched to experimental constraints and noise models.

% ============================================================
% References
% ============================================================
\bibliography{libraryB}

% =====================================================================
% Supplemental Material
% =====================================================================
\clearpage
\onecolumngrid
\setcounter{equation}{0}
\setcounter{figure}{0}
\setcounter{table}{0}
\setcounter{section}{0}
\renewcommand{\theequation}{S\arabic{equation}}
\renewcommand{\thefigure}{S\arabic{figure}}
\renewcommand{\thesection}{S-\Roman{section}}

% Fix for hyperref duplicate destinations when counters are reset
\renewcommand{\theHequation}{SM.\arabic{equation}}
\renewcommand{\theHfigure}{SM.\arabic{figure}}
\renewcommand{\theHsection}{SM.\Roman{section}}

% =====================================================================
% SUPPLEMENTARY MATERIAL (REPLACEMENT DRAFT)
%=====================================================================
\section{SUPPLEMENTARY MATERIAL}
\section{Petz \texorpdfstring{$f$}{f}-divergences in left/right multiplication form}

Let $\mathcal{H}$ be finite-dimensional. For a positive definite operator $\tau>0$, define the left- and right-multiplication superoperators on $\mathcal{B}(\mathcal{H})$ by
\begin{equation}
\label{SM:eq:LR_def}
L_\tau(X):=\tau X,
\qquad
R_\tau(X):=X\tau .
\end{equation}
For states $\rho,\sigma>0$, define the (relative) modular operator
\begin{equation}
\label{SM:eq:Delta_def}
\Delta_{\rho,\sigma}:=L_\rho R_\sigma^{-1}.
\end{equation}
Let $f:(0,\infty)\to\mathbb{R}$ be operator convex with $f(1)=0$. The Petz $f$-divergence (quasi-entropy) can be written as
\begin{equation}
\label{SM:eq:Petz_f_div}
D_f(\rho\|\sigma)
:=
\mathrm{Tr}\!\left[\sigma^{1/2}\, f(\Delta_{\rho,\sigma})\big(\sigma^{1/2}\big)\right]
=
\left\langle \sigma^{1/2},\, f(\Delta_{\rho,\sigma})\big(\sigma^{1/2}\big)\right\rangle_{\mathrm{HS}} ,
\end{equation}
where $\langle A,B\rangle_{\mathrm{HS}}:=\mathrm{Tr}(A^\dagger B)$.

The special case $f(t)=t\log t$ yields the quantum relative entropy
\begin{equation}
\label{SM:eq:KL_def}
D(\rho\|\sigma):=\mathrm{Tr}\!\left[\rho(\log\rho-\log\sigma)\right].
\end{equation}

\section{One collision step, entropy production, mutual information split, and bath relative entropy}

Consider a single collision with initial product state $\rho_S\otimes\rho_E$ and unitary $U$ on $S\otimes E$:
\begin{equation}
\label{SM:eq:collision_map}
\rho'_{SE}=U(\rho_S\otimes\rho_E)U^\dagger,
\qquad
\rho'_S=\mathrm{Tr}_E(\rho'_{SE}),
\qquad
\rho'_E=\mathrm{Tr}_S(\rho'_{SE}).
\end{equation}
We define the (KL) entropy production as the relative entropy to the nonequilibrium reference $\rho'_S\otimes\rho_E$,
\begin{equation}
\label{SM:eq:Sigma_KL_def}
\Sigma
:=
D\!\left(\rho'_{SE}\,\middle\|\,\rho'_S\otimes\rho_E\right).
\end{equation}
Using standard identities for relative entropy of bipartite states, $\Sigma$ splits into a nonnegative mutual-information term and the bath relative entropy:
\begin{equation}
\label{SM:eq:Sigma_split}
\Sigma
=
I(S{:}E)_{\rho'_{SE}} + D(\rho'_E\|\rho_E),
\qquad
I(S{:}E)_{\rho'_{SE}}:=D\!\left(\rho'_{SE}\,\middle\|\,\rho'_S\otimes\rho'_E\right)\ge 0 .
\end{equation}
In the numerical plots labeled ``Dbath'', we focus on the bath relative entropy alone,
\begin{equation}
\label{SM:eq:Dbath_def}
\Dbath:=D(\rho'_E\|\rho_E),
\end{equation}
which removes the (positive) mutual information contribution and is therefore a sharper target for saturation.

\section{Quadratic contrasts: \texorpdfstring{$\chi^2_\lambda$}{chi2-lambda} and integral representations}

For $\lambda\in(0,1)$ and $\sigma>0$, define the super operator 
\begin{equation}
\label{SM:eq:K_lambda_def}
K_{\lambda,\rho,\sigma}:=(1-\lambda)L_\rho+\lambda R_\sigma .
\end{equation}
The $\chi^2_\lambda$ divergence is the quadratic form
\begin{equation}
\label{SM:eq:chi2_lambda_def}
\chi^2_\lambda(\rho\|\sigma)
:=
\mathrm{Tr}\!\left[(\rho-\sigma)\,K_{\lambda,\rho,\sigma}^{-1}(\rho-\sigma)\right].
\end{equation}

A key input is that operator-convex $f$-divergences admit a decomposition into $\chi^2_\lambda$ modes: there exists a nonnegative weight function $w_f(\lambda)$ (depending on $f$) such that
\begin{equation}
\label{SM:eq:f_integral_rep}
D_f(\rho\|\sigma)
=
\int_0^1 w_f(\lambda)\,\chi^2_\lambda(\rho\|\sigma)\,d\lambda .
\end{equation}
For KL, the weight is $w_{\mathrm{KL}}(\lambda)=\lambda$, giving
\begin{equation}
\label{SM:eq:KL_integral_rep}
D(\rho\|\sigma)
=
\int_0^1 \lambda\,\chi^2_\lambda(\rho\|\sigma)\,d\lambda .
\end{equation}

\section{Currents, covariance matrices, and a one-dimensional witness}

Let $\vec Q=(Q_1,\dots,Q_m)$ be a list of (possibly noncommuting) reservoir charges on $E$. Define the one-step current vector
\begin{equation}
\label{SM:eq:current_vector_def}
\Delta \vec q
:=
\Big(\mathrm{Tr}\big[(\rho'_E-\rho_E)Q_1\big],\dots,\mathrm{Tr}\big[(\rho'_E-\rho_E)Q_m\big]\Big)\in\mathbb{R}^m .
\end{equation}
For a real vector $u\in\mathbb{R}^m$, define the scalar charge witness and its transported current
\begin{equation}
\label{SM:eq:witness_def}
Q_u:=u\cdot\vec Q=\sum_{i=1}^m u_i Q_i,
\qquad
x(u):=\mathrm{Tr}\big[(\rho'_E-\rho_E)Q_u\big]=u\cdot\Delta\vec q .
\end{equation}

We use the symmetrized covariance matrix in state $\tau$,
\begin{equation}
\label{SM:eq:cov_def}
\mathrm{Cov}_\tau(Q_i,Q_j)
:=
\frac12\,\mathrm{Tr}\!\left[\tau\,\{\Delta_\tau Q_i,\Delta_\tau Q_j\}\right],
\qquad
\Delta_\tau Q_i:=Q_i-\mathrm{Tr}(\tau Q_i)\,\mathbb{I}.
\end{equation}
Define the pre- and post-collision covariance matrices
\begin{equation}
\label{SM:eq:V_Vp_def}
V_{ij}:=\mathrm{Cov}_{\rho_E}(Q_i,Q_j),
\qquad
V'_{ij}:=\mathrm{Cov}_{\rho'_E}(Q_i,Q_j).
\end{equation}
Then $\mathrm{Var}_{\rho_E}(Q_u)=u^\top V u$ and $\mathrm{Var}_{\rho'_E}(Q_u)=u^\top V' u$.

\section{TUR from Chapman--Robbins at fixed \texorpdfstring{$\lambda$}{lambda}: general \texorpdfstring{$f$}{f} and KL corollary}

For $\lambda\in(0,1)$ define the tight Chapman--Robbins contrast
\begin{equation}
\label{SM:eq:CR_h_def}
h_\lambda(x,y,z)
:=
\frac{x^2}{(1-\lambda)y+\lambda z+\lambda(1-\lambda)x^2}.
\end{equation}

\subsection{Theorem: scalar TUR for a witness observable}

\begin{theorem}[Witness TUR for $D_f(\rho'_E\|\rho_E)$]
\label{SM:thm:witness_TUR_f}
For any operator-convex $f$ with integral weight $w_f(\lambda)\ge 0$ as in \eqref{SM:eq:f_integral_rep}, and any $u\in\mathbb{R}^m$,
\begin{equation}
\label{SM:eq:witness_TUR_f}
D_f(\rho'_E\|\rho_E)
\ge
\int_0^1 w_f(\lambda)\,
h_\lambda\!\Big(x(u),\,\mathrm{Var}_{\rho'_E}(Q_u),\,\mathrm{Var}_{\rho_E}(Q_u)\Big)\,d\lambda .
\end{equation}
\end{theorem}

\paragraph{Proof.}
Apply \eqref{SM:eq:f_integral_rep} with $(\rho,\sigma)=(\rho'_E,\rho_E)$, and use the tight Chapman--Robbins lower bound for each $\chi^2_\lambda(\rho'_E\|\rho_E)$ with witness $Q_u$, which yields
\begin{equation}
\label{SM:eq:chi2_CR_step}
\chi^2_\lambda(\rho'_E\|\rho_E)
\ge
h_\lambda\!\Big(x(u),\,\mathrm{Var}_{\rho'_E}(Q_u),\,\mathrm{Var}_{\rho_E}(Q_u)\Big).
\end{equation}
Integrating \eqref{SM:eq:chi2_CR_step} against $w_f(\lambda)\,d\lambda$ gives \eqref{SM:eq:witness_TUR_f}.
\qed

\begin{corollary}[KL witness bound]
\label{SM:cor:KL_witness}
For KL, $w_{\mathrm{KL}}(\lambda)=\lambda$ and therefore
\begin{equation}
\label{SM:eq:witness_TUR_KL}
\Dbath=D(\rho'_E\|\rho_E)
\ge
\int_0^1 \lambda\,
\frac{(u\cdot\Delta\vec q)^2}{(1-\lambda)\,u^\top V' u+\lambda\,u^\top V u+\lambda(1-\lambda)(u\cdot\Delta\vec q)^2}\,d\lambda .
\end{equation}
\end{corollary}

\subsection{Theorem: matrix TUR via optimizing over \texorpdfstring{$u$}{u}}

For each $\lambda\in(0,1)$ define the convex combination and the curvature factor
\begin{equation}
\label{SM:eq:Mlambda_alpha_def}
M_\lambda:=(1-\lambda)V' + \lambda V,
\qquad
\alpha_\lambda:=\lambda(1-\lambda).
\end{equation}
Assume $M_\lambda$ is invertible on the support of $\Delta\vec q$. Define the scalar
\begin{equation}
\label{SM:eq:s_lambda_def_SM}
s_\lambda:=\Delta\vec q^\top M_\lambda^{-1}\Delta\vec q .
\end{equation}

\begin{lemma}[Optimization identity at fixed $\lambda$]
\label{SM:lem:opt_identity}
For fixed $\lambda\in(0,1)$,
\begin{equation}
\label{SM:eq:opt_identity}
\sup_{u\neq 0}\;
\frac{(u\cdot\Delta\vec q)^2}{u^\top M_\lambda u+\alpha_\lambda (u\cdot\Delta\vec q)^2}
=
\frac{s_\lambda}{1+\alpha_\lambda s_\lambda},
\end{equation}
and an optimizer is $u\propto M_\lambda^{-1}\Delta\vec q$.
\end{lemma}

\paragraph{Proof.}
Write the denominator as $u^\top M_\lambda u+\alpha_\lambda (u^\top \Delta\vec q)^2$.
Let $v:=M_\lambda^{1/2}u$ and $a:=M_\lambda^{-1/2}\Delta\vec q$. Then
\begin{equation}
\label{SM:eq:opt_proof_change}
\frac{(u^\top \Delta\vec q)^2}{u^\top M_\lambda u+\alpha_\lambda (u^\top \Delta\vec q)^2}
=
\frac{(v^\top a)^2}{\|v\|_2^2+\alpha_\lambda (v^\top a)^2}.
\end{equation}
For fixed $a$, the quantity $(v^\top a)^2/\|v\|_2^2$ is maximized by $v\parallel a$ (Cauchy--Schwarz), with maximum $\|a\|_2^2=a^\top a=\Delta\vec q^\top M_\lambda^{-1}\Delta\vec q=s_\lambda$. Substituting yields \eqref{SM:eq:opt_identity}. The maximizing direction $v\parallel a$ corresponds to $u\propto M_\lambda^{-1}\Delta\vec q$.
\qed

\begin{theorem}[Matrix TUR for the bath relative entropy]
\label{SM:thm:matrix_TUR_KL}
The KL bath relative entropy satisfies the fully optimized integral bound
\begin{equation}
\label{SM:eq:matrix_TUR_KL}
\Dbath
\ge
\int_0^1 \lambda\,
\frac{s_\lambda}{1+\lambda(1-\lambda)s_\lambda}\,d\lambda,
\qquad
s_\lambda=\Delta\vec q^\top\big((1-\lambda)V'+\lambda V\big)^{-1}\Delta\vec q .
\end{equation}
\end{theorem}

\paragraph{Proof.}
Start from the KL witness bound \eqref{SM:eq:witness_TUR_KL}. For each fixed $\lambda$, the integrand is exactly the objective in Lemma~\ref{SM:lem:opt_identity} with $M_\lambda$ and $\alpha_\lambda$. Optimizing over $u$ at fixed $\lambda$ yields $\frac{s_\lambda}{1+\alpha_\lambda s_\lambda}$, and integrating over $\lambda$ gives \eqref{SM:eq:matrix_TUR_KL}.
\qed

{\it Remark on Quantum f-Divergences}
Our main result can also be stated for general Quantum Petz f-divergences. Using the integral representation, 
\begin{equation}
\label{eq:formalism_f_chi2}
D_{\mathrm{f}}(\rho'_E\|\rho_E)=\int_0^1 d\lambda\;w_{\mathrm{f}}(\lambda)
\chi^2_\lambda(\rho'_E\|\rho_E),
\end{equation}
for positive weights $w_f(\lambda)$. It results in
\begin{equation}
D_{\mathrm{f}}(\rho_E'\|\rho_E)\geq B_{\mathrm{f}}(\Delta q,V,V'),
\end{equation}
where $B_{\mathrm{f}}(\Delta q,V,V')=\int_0^1 d\lambda\, w_{\mathrm{f}}(\lambda)s_\lambda/(1+\lambda(1-\lambda)s_\lambda)$. The entropy production is the specific Kullback-Leibler case $w_{KL}(\lambda)=\lambda$. Another notable example is the affinity/Bures–Hellinger family $H^2(\rho,\sigma):=1-\Tr(\sqrt\rho \sqrt{\sigma})$, where $w_{\mathrm{f}}(\lambda)=(1/\pi)\sqrt{\lambda(1-\lambda)}$.

\section{Small-current limit: quadratic form and relation to Onsager geometry}

In a near-fixed-point regime, $\rho'_E=\rho_E+O(\varepsilon)$, implying
\begin{equation}
\label{SM:eq:near_fp_scaling}
V' = V + O(\varepsilon),
\qquad
\Delta\vec q = O(\varepsilon).
\end{equation}
Then $s_\lambda=s+O(\varepsilon^3)$ with
\begin{equation}
\label{SM:eq:s_simple_def_SM}
s:=\Delta\vec q^\top V^{-1}\Delta\vec q .
\end{equation}
Substituting $V'=V$ into \eqref{SM:eq:matrix_TUR_KL} yields a closed form
\begin{equation}
\label{SM:eq:F_s_def_SM}
\Dbath
\ge
F(s),
\qquad
F(s):=\int_0^1 \lambda\,\frac{s}{1+\lambda(1-\lambda)s}\,d\lambda
=
2\sqrt{\frac{s}{s+4}}\;\mathrm{artanh}\!\sqrt{\frac{s}{s+4}}.
\end{equation}
Expanding at small $s$ gives
\begin{equation}
\label{SM:eq:F_small_s}
F(s)=\frac{s}{2}-\frac{s^2}{12}+O(s^3),
\qquad\Longrightarrow\qquad
\Dbath
\ge
\frac12\,\Delta\vec q^\top V^{-1}\Delta\vec q + O(\|\Delta\vec q\|^4).
\end{equation}

It is useful to relate this to the matrix used in standard linear-response notation. Define
\begin{equation}
\label{SM:eq:Ceff_def}
(C_{\mathrm{eff}})_{ij}:=\mathrm{Tr}\!\left[\rho_E\,\{\Delta_{\rho_E}Q_i,\Delta_{\rho_E}Q_j\}\right].
\end{equation}
By \eqref{SM:eq:cov_def}--\eqref{SM:eq:V_Vp_def}, we have $C_{\mathrm{eff}}=2V$, hence the leading quadratic bound can also be written as
\begin{equation}
\label{SM:eq:quadratic_bound_Ceff}
\Dbath
\ge
\Delta\vec q^\top C_{\mathrm{eff}}^{-1}\Delta\vec q + O(\|\Delta\vec q\|^4).
\end{equation}

\paragraph{Comparison with Kubo--Mori.}
The Hessian of $D(\rho_E+\delta\rho\|\rho_E)$ at $\delta\rho=0$ defines the Kubo--Mori (Bogoliubov) inner product and its associated covariance matrix $V^{\mathrm{KM}}$ (a linear-response object). Replacing $V^{\mathrm{KM}}$ by the symmetrized covariance $V$ typically yields a looser quadratic bound, but the advantage is that $V$ is directly a noise (fluctuation) matrix and is exactly what is estimated in the simulation.

% ============================================================
\section*{Applications: strong-coupling qubit collision model (numerical experiment)}
% ============================================================

\paragraph{Goal.}
We numerically demonstrate that the bath relative entropy
$\Dbath:=D(\rho'_E\|\rho_E)$ generated by a \emph{single strong-coupling collision}
satisfies the matrix TUR lower bound
$\Dbath\ge B(\Delta q,V,V')$ from Eq.~\eqref{eq:formalism_main_chain},
and we map out when the bound becomes nearly saturated.
The simulation implements the microscopic definitions in
Eqs.~\eqref{eq:EP_def}--\eqref{eq:EP_split} and the bound in
Eqs.~\eqref{eq:s_lambda_intro}--\eqref{eq:formalism_main_chain}.

\subsection*{Microscopic setup: one-shot collision at strong coupling}

\paragraph{Systems.}
Both the system $S$ and bath probe $E$ are qubits ($\dim S=\dim E=2$).
A \emph{single collision} starts from a product input
\begin{equation}
\rho_{SE}=\rho_S\otimes\rho_E,
\end{equation}
and evolves under a joint unitary $U_{SE}$ to
\begin{equation}
\rho'_{SE}=U_{SE}(\rho_S\otimes\rho_E)U_{SE}^\dagger,
\qquad
\rho'_S=\Tr_E\rho'_{SE},
\qquad
\rho'_E=\Tr_S\rho'_{SE}.
\end{equation}
The entropy production $\Sigma$ is computed from the process-level definition
(Eq.~\eqref{eq:EP_def}) and obeys $\Sigma\ge \Dbath$ (Eq.~\eqref{eq:EP_split}).

\paragraph{Why this is \emph{strong coupling}.}
We do \emph{not} assume weak interaction, a Markovian generator, or a perturbative GKLS limit.
Instead, we directly apply a generic two-qubit unitary $U_{SE}$ with a tunable interaction angle
$\phi$ that can be $O(1)$ (including values close to $\pi/2$), so a \emph{single collision}
can create large correlations $I(S\!:\!E)_{\rho'}$ and non-negligible bath relative entropy
$D(\rho'_E\|\rho_E)$.
This is the nonperturbative regime in which linear-response approximations are not expected to hold
a priori, and where saturation/non-saturation of the fully nonlinear integrand in
Eq.~\eqref{eq:formalism_main_chain} is genuinely informative.

\subsection*{State preparation: generic bath state and controlled system deviation}

\paragraph{Bath reference state $\rho_E$.}
We sample $\rho_E$ as a generic full-rank qubit state via its Bloch representation
\begin{equation}
\rho_E=\frac{1}{2}\Big(\I + r\, \hat n\cdot \vec\sigma\Big),
\qquad
0<r<1,\ \ \hat n\in S^2,
\end{equation}
where $\vec\sigma=(\sigma_x,\sigma_y,\sigma_z)$.
The Bloch direction $\hat n$ is Haar-uniform on the sphere, and the radius $r$ is drawn from a
mixture distribution designed to cover both moderately mixed and near-pure regimes, subject to
configurable cutoffs
\begin{equation}
r\in[r_{\min},r_{\max}],\qquad r_{\min}=0.1,\ \ r_{\max}=0.95 \ \ (\text{default}).
\end{equation}
This produces a broad range of bath purities $\Tr(\rho_E^2)=(1+r^2)/2$ and eigenvalue ratios, which
stress-tests numerical stability of $D(\rho'_E\|\rho_E)$ and the covariance inversions below.

\paragraph{System input state $\rho_S$.}
To probe both near-equilibrium and far-from-equilibrium behavior while retaining a clean control
parameter, we sample $\rho_S$ \emph{relative to} $\rho_E$ using three modes:
\begin{itemize}
\item \emph{Haar isospectral (default):}
$\rho_S=U\rho_EU^\dagger$ with $U\in\mathrm{SU}(2)$ Haar-random.
This keeps $\rho_S$ and $\rho_E$ isospectral (same eigenvalues) but generically misaligned
(equivalently, it injects coherence relative to the bath eigenbasis).
\item \emph{Small isospectral (near-fixed-point control):}
$\rho_S=U(\varepsilon)\rho_EU(\varepsilon)^\dagger$ where
$U(\varepsilon)=e^{-i\varepsilon(\hat a\cdot \vec\sigma)/2}$ rotates around a random axis $\hat a$,
with $\varepsilon\in[\varepsilon_{\min},\varepsilon_{\max}]$ drawn log-uniformly.
This targets the regime where $V'\approx V$ and the Onsager/small-fluctuation expansion becomes visible.
\item \emph{Independent random:}
$\rho_S$ is an independent Bloch-sampled qubit state, to explore fully generic nonequilibrium inputs.
\end{itemize}

\subsection*{Interaction unitary: partial swap + bath-aware fixed-point block}

\paragraph{Two knobs for strong-coupling mixing.}
We use a family of two-qubit unitaries with two independent sources of noncommuting dynamics.

\paragraph{(i) Partial swap.}
Define the SWAP operator on $S\otimes E$ by $\mathrm{SWAP}\,\ket{ij}=\ket{ji}$.
The partial-swap unitary is
\begin{equation}
U_{\mathrm{swap}}(\phi)
:=e^{-i\phi\,\mathrm{SWAP}}
=\cos(\phi)\,\I_{SE}-i\sin(\phi)\,\mathrm{SWAP},
\qquad
\phi\in[\phi_{\min},\phi_{\max}],
\end{equation}
with default sampling range $\phi\in[0.05,1.57]$.
This is a canonical collision-model interaction: it continuously interpolates between identity and
(full) SWAP, and can strongly exchange information/charges between $S$ and $E$ in a single shot.

\paragraph{(ii) $\rho_E$-aware fixed-point unitary.}
We additionally include a \emph{bath-adapted} unitary $U_{\mathrm{fp}}$ constructed in the eigenbasis of $\rho_E$.
Let $\rho_E=W\mathrm{diag}(p_0,p_1)W^\dagger$.  In the basis where $\rho_E\otimes\rho_E$ is diagonal, define
\begin{equation}
U_{\mathrm{fp}}^{(\mathrm{eig})}
=
\begin{pmatrix}
e^{i\alpha} & 0 & 0 & 0\\
0 & & & 0\\
0 & & U_{2\times2} & 0\\
0 & 0 & 0 & e^{i\beta}
\end{pmatrix},
\qquad
\alpha,\beta\sim\mathrm{Unif}[0,2\pi),\ \ U_{2\times2}\sim\mathrm{Haar\ on}\ U(2),
\end{equation}
acting as arbitrary mixing in the $\{\ket{01},\ket{10}\}$ subspace while only phasing $\ket{00}$ and $\ket{11}$.
Transforming back,
\begin{equation}
U_{\mathrm{fp}}=(W\otimes W)\,U_{\mathrm{fp}}^{(\mathrm{eig})}\,(W\otimes W)^\dagger,
\end{equation}
which \emph{exactly preserves} $\rho_E\otimes\rho_E$ as a fixed point, i.e.,
$U_{\mathrm{fp}}(\rho_E\otimes\rho_E)U_{\mathrm{fp}}^\dagger=\rho_E\otimes\rho_E$.

\paragraph{(iii) Composed strong-coupling unitary (default).}
Our default interaction is the composition
\begin{equation}
U_{SE}=U_{\mathrm{swap}}(\phi)\,U_{\mathrm{fp}},
\end{equation}
which provides two independent noncommuting ``knobs'':
the swap angle $\phi$ controlling direct exchange strength, and the bath-eigenbasis block
controlling additional coherent mixing that is \emph{adapted to the chosen} $\rho_E$.

\subsection*{Currents and non-Abelian charge geometry (what we measure)}

\paragraph{Charge observables.}
We consider $m=k=2$ charge observables $q=(q_u)_u$ represented by qubit operators
\begin{equation}
q_u \equiv Q_u=\frac{1}{2}\,\hat e_u\cdot \vec\sigma.
\end{equation}
To avoid privileged axes, we often sample a \emph{random charge frame} by drawing a random rotation
$R\in\mathrm{SO}(3)$ and setting $\{\hat e_u\}$ to be (a subset of) the rotated orthonormal axes:
\begin{itemize}
\item $k=2$: choose $\hat e_1=R\hat x$ and $\hat e_2=R\hat z$ (two noncommuting charges).
\end{itemize}
This implements a genuinely non-Abelian setting at the level of observables: for $k\ge2$ the $Q_u$
do not commute and define a matrix-valued covariance geometry.

\paragraph{Currents.}
In the collision model, one can evaluate transported charge either on $S$ or on $E$ (depending on the
operational convention).  In the numerics we use the bath-side change
\begin{equation}
\Delta q_u
:=\Tr\!\big[(\rho'_E-\rho_E)\,Q_u\big],
\qquad
\Delta q=\big(\Delta q_u\big)_{u=1}^k\in\mathbb{R}^k,
\end{equation}
which is directly computable from $\rho'_E$ and has the same status as the current vector
introduced in Eq.~\eqref{eq:Deltaq_def}.

\paragraph{Covariances.}
We compute the (real, symmetric) covariance matrix of the charges in the initial and final bath states:
\begin{equation}
V_{uv}:=\frac{1}{2}\Tr\!\left[\rho_E\{\Delta Q_u,\Delta Q_v\}\right],
\qquad
V'_{uv}:=\frac{1}{2}\Tr\!\left[\rho'_E\{\Delta Q'_u,\Delta Q'_v\}\right],
\end{equation}
where $\Delta Q_u=Q_u-\Tr(\rho_E Q_u)\,\I$ and similarly for $\Delta Q'_u$ with $\rho'_E$.
These are the $V$ and $V'$ entering Eq.~\eqref{eq:s_lambda_intro}.

\subsection*{Step-by-step simulation pipeline}

For each Monte Carlo sample, we perform:

\begin{enumerate}
\item \textbf{Sample the bath state.}
Draw $(r,\hat n)$ and set $\rho_E=\frac12(\I+r\,\hat n\cdot\vec\sigma)$.

\item \textbf{Choose the charge vector.}
Either fix $\{Q_u\}$ to Pauli directions (legacy) or sample a random frame $R$ and define
$Q_u=\frac12 (R\hat e_u)\cdot\vec\sigma$ for $k=2$.

\item \textbf{Sample the system state.}
Construct $\rho_S$ using one of the modes above (typically Haar-isospectral relative to $\rho_E$).

\item \textbf{Sample the strong-coupling unitary.}
Draw $\phi\in[\phi_{\min},\phi_{\max}]$, and draw $(\alpha,\beta,U_{2\times2})$ to build $U_{\mathrm{fp}}$,
then set $U_{SE}=U_{\mathrm{swap}}(\phi)U_{\mathrm{fp}}$.

\item \textbf{Evolve and reduce.}
Compute
$\rho'_{SE}=U_{SE}(\rho_S\otimes\rho_E)U_{SE}^\dagger$ and partial traces
$\rho'_S=\Tr_E\rho'_{SE}$, $\rho'_E=\Tr_S\rho'_{SE}$.

\item \textbf{Compute EP and bath relative entropy.}
Compute
\begin{equation}
\Sigma = D\!\left(\rho'_{SE}\,\big\|\,\rho'_S\otimes\rho_E\right),
\qquad
\Dbath=D(\rho'_E\|\rho_E),
\end{equation}
using spectral matrix logarithms (eigendecomposition with a small eigenvalue floor for stability).

\item \textbf{Compute currents and covariances.}
Compute $\Delta q_u=\Tr[(\rho'_E-\rho_E)Q_u]$ and the matrices $V,V'$ as above.
We also record diagnostics such as the relative covariance drift
$\|V'-V\|/\|V\|$ (using Frobenius).

\item \textbf{Evaluate the bound.}
For $\lambda\in(0,1)$, define the interpolated covariance $V_\lambda$ and the scalar
\begin{equation}
s_\lambda=\Delta q^{\mathsf T}V_\lambda^{-1}\Delta q,
\end{equation}
and compute the integral in Eq.~\eqref{eq:formalism_main_chain} by Gauss--Legendre quadrature on $[0,1]$.
Numerically, we regularize the inversion when $V_\lambda$ becomes ill-conditioned.

\subsection*{Sampling strategies: how we find saturation}

A purely uniform Monte Carlo can miss the near-saturation ``manifold'' when it occupies a small
volume in parameter space.  We therefore use three strategies:
\begin{itemize}
\item \emph{Monte Carlo:} unbiased baseline over all parameters.
\item \emph{Stratified sampling:} bin points by a chosen x-axis variable (e.g.\ $s$ or the full bound)
to ensure uniform coverage across decades.
\item \emph{Saturation hunt:} start from a stratified baseline and then preferentially keep/refine the
smallest-gap samples per bin, yielding clouds that visibly approach the diagonal
$\Dbath=B(\Delta q,V,V')$ in the master plot.
\end{itemize}
\end{enumerate}
\subsection*{Small current regime and the inset theory curve}

When $V'\approx V$ (small covariance drift, typical of the small-isospectral mode), the full integral bound
collapses to a one-parameter function of the simple signal-to-noise ratio
$s:=\Delta q^{\mathsf T}V^{-1}\Delta q$,
\begin{equation}
\Dbath \approx F(s),
\qquad
F(s)=2\sqrt{\frac{s}{s+4}}\;\mathrm{artanh}\!\sqrt{\frac{s}{s+4}},
\end{equation}
with small-$s$ expansion $F(s)=\frac{s}{2}-\frac{s^2}{12}+O(s^3)$.
We overlay $F(s)$ and its quadratic truncation in the inset to show explicitly how the
Onsager-type quadratic form emerges as the small-fluctuation limit while the main panel
displays the fully nonlinear strong-coupling behavior.

\end{document}